\documentclass[preprint,aps,tightenlines,floatfix,showpacs,byrevtex]{revtex4}
\usepackage{epsfig}
\usepackage{graphicx}

\newcommand{\del}{\partial}
\newcommand{\dmu}{\partial_\mu}
\newcommand{\dnu}{\partial_\nu}
\newcommand{\Amu}{A_\mu}
\newcommand{\Anu}{A_\nu}

\newcommand{\rf}[4]{{\em {#1}} {\bf #2}, #3 (#4)}

\newcommand{\beq}{\begin{equation}}
\newcommand{\eeq}{\end{equation}}
\newcommand{\beqa}{\begin{eqnarray}}
\newcommand{\eeqa}{\end{eqnarray}}

\newcommand{\cO}{\cal O}

\newbox\rotbox

\begin{document}

\preprint{\vbox{
\rightline{ADP-01-50/T482}}}

\title{Highly-improved lattice field-strength tensor}

\author{Sundance O.\ Bilson-Thompson\footnote{
E-mail:~sbilson@physics.adelaide.edu.au}}
\author{Derek B.\ Leinweber\footnote{
E-mail:~dleinweb@physics.adelaide.edu.au ~$\bullet$~ Tel:
+61~8~8303~3423 ~$\bullet$~ Fax: +61~8~8303~3551 \hfill\break
\null\quad\quad
WWW:~http://www.physics.adelaide.edu.au/theory/staff/leinweber/}}
\author{Anthony G.\ Williams\footnote{
E-mail:~awilliam@physics.adelaide.edu.au ~$\bullet$~ Tel:
+61~8~8303~3546 ~$\bullet$~ Fax: +61~8~8303~3551 \hfill\break
\null\quad\quad
WWW:~http://www.physics.adelaide.edu.au/theory/staff/williams.html}}
\affiliation{CSSM Lattice Collaboration, Special Research Centre for the 
Subatomic Structure of Matter and Department of Physics and Mathematical 
Physics, University of Adelaide, Adelaide SA 5005, Australia}
\date{\today}

\begin{abstract}
We derive an ${\cal O}(a^4)$-improved lattice version of the continuum 
field-strength tensor. Discretization errors are reduced via the 
combination of several clover terms of various sizes, complemented by 
tadpole improvement. The resulting improved field-strength tensor is used 
to construct ${\cal O}(a^4)$-improved topological charge and action operators. 
We compare the values attained by these operators as we cool several 
configurations to self-duality with a previously defined highly-improved 
action and assess the relative scale of the remaining discretization errors.
\end{abstract}

\vspace{1cm}
 \parskip=2mm

\maketitle



\section{Introduction}
\label{sec:intro}
Lattice gauge theory has shown itself to be an extremely useful tool for studying 
non-perturbative physics. Its success is founded on the ability to systematically 
remove errors introduced by the discretization of space-time. The magnitude of such 
errors are determined by the lattice spacing $a$. In the continuum limit 
$a \rightarrow 0$ these error terms vanish, and so one could expect to asymptotically 
approach continuum physics by moving to finer lattice spacings. Unfortunately the 
computational cost increases significantly as the spacing decreases.\\

It is advantageous to `improve' operators by identifying and algebraically eliminating 
discretization errors from their lattice definitions \cite{Syman}.  In 
particular, calculations of the Yang-Mills action based upon the strategic combination of 
several different Wilson loop terms and the reduction of non-classical errors by tadpole 
improvement \cite{LePage} have demonstrated a significant decrease in deviations from 
expected continuum physics in smoothing algorithms such as cooling and smearing (see for 
instance Ref.~\cite{Adel1} and references therein). In this paper we consider the construction of 
${\cO}(a^4)$-improved gluon field-strength tensors, $F_{\mu\nu}$, from which both the topological
charge density and the action may be constructed. In the following we refer to such an 
action as the ``reconstructed action''.\\
 
The accuracy of our improved operators is investigated by calculating the action, 
the reconstructed action, and the topological charge on an ensemble of field 
configurations. We then cool these rough configurations to produce highly self-dual 
configurations with a range of topological charges. Thermalization and cooling are 
carried out by use of parallel algorithms with appropriate link partitioning \cite{AdelPar}.
Comparisons of the topological charge $Q$, and action normalized to the single-instanton 
action, $S/S_0$, provide quantitative measures of the merit of each of the improvement 
schemes considered.\\
 
This paper is set out as follows: In Section~\ref{sec:LatActionandQ} we look at the 
lattice definitions of the action and topological charge operators and give 
a brief discussion of the construction of the ${\cal O}(a^4)$-improved forms of 
these operators \cite{DeForc}. In Section~\ref{sec:LatFmunu} we proceed to describe the 
construction of an analogously improved lattice version of the field-strength tensor, 
$F_{\mu\nu}$. This tensor is used to construct an improved topological charge operator 
and the reconstructed action. In Section~\ref{sec:Analysis} we use the cooling of 
gauge-fields on the lattice to determine which form of improvement gives the most 
continuum-like results, and compare the standard improved action with the reconstructed 
action to verify consistency. Our results and conclusions are presented in 
Section~\ref{sec:Discussions}.

\section{Lattice Action and Topological Charge}
\label{sec:LatActionandQ}
\subsection{Wilson Action}
The lattice version of the Yang-Mills action was first proposed by Wilson \cite{Wilson}. 
The action is calculated from the plaquette, a closed product of four link operators 
incorporating the link $U_{\mu}$, 
\begin{equation}
S_{\rm {Wil}} = \beta \sum_{x} \sum_{\mu<\nu}
	        \left(1-\frac{1}{N}Re{\rm {Tr}} W^{(1\times 1)}_{\mu\nu}\right)\, ,
\label{eq:WilsonAction}
\end{equation}
for an $SU(N)$ field theory where the plaquette operator $W^{(1\times 1)}_{\mu\nu}$ is 
\begin{equation}
W^{(1\times 1)}_{\mu\nu} = U_{\mu}(x)U_{\nu}(x+a\hat{\mu})U^{\dag}_{\mu}(x+a\hat{\nu})U^{\dag}_{\nu}(x)\, .
\label{eq:plaquette}
\end{equation}
The link variables $U_\mu(x)$ are defined by the Schwinger line integral
\beq
U_\mu(x) = \ {\cal P} \exp\left\{ig \int_x^{x+a\mu} dz A_{\mu}(z)\right\}\, ,
\label{eq:Schwinger}
\eeq
and are in general non-Abelian (hence the ${\cal P}$ on the right-hand side of Eq.~(\ref{eq:Schwinger})
to denote path-ordering). 
Let us define $W^{(m \times n)}_{\mu\nu}$ to be the closed loop product in the $\mu-\nu$ plane
with extent $m$ lattice spacings in the $\mu$-direction and $n$ lattice spacings in the 
$\nu$-direction. Therefore $W^{(1\times 1)}_{\mu\nu}$ is the usual plaquette, $W^{(1\times 2)}_{\mu\nu}$ is an
$(a\times 2a)$ loop
\beq
W^{(1\times2)}_{\mu\nu}=
      U_{\mu}(x)U_{\nu}(x+a\hat{\mu})U_{\nu}(x+a\hat{\mu}+a\hat{\nu})U^{\dag}_{\mu}(x+2a\hat{\nu})
      U^{\dag}_{\nu}(x+a\hat{\nu})U^{\dag}_{\nu}(x)\, ,
\eeq
and so on. The Wilson action based on the plaquette for QCD 
(i.e., $N=3$) expanded around $x_0$, the centre of the plaquette, may be written as follows
\begin{eqnarray}
S_{\rm {Wil}} & =  {\displaystyle \beta \sum_{x_0} \sum_{\mu<\nu} } & \left[
	1 - \frac{1}{N}Re{\rm {Tr}}W_{\mu \nu}^{(1 \times 1)}(x_0) \right], \nonumber \\
	& =  {\displaystyle \beta \sum_{x_0} \sum_{\mu<\nu} } & \left[
	1 - 1 + \frac{a^4 g^2}{6} {\rm {Tr}} F_{\mu \nu}^2 (x_0) + {\cal O}(g^2a^6)+ {\cal O}(g^4a^6) \right], \nonumber \\
	& =  {\displaystyle \beta \sum_{x_0} \sum_{\mu<\nu} } & \left[
        \frac{a^4 g^2}{6} {\rm {Tr}} F_{\mu \nu}^2 (x_0) +{\cal O}(g^2a^6)+ {\cal O}(g^4a^6) \right], \nonumber \\
	& =  {\displaystyle a^4 \sum_{x_0} \sum_{\mu<\nu} } & \left[
        {\rm {Tr}} F_{\mu \nu}^2 (x_0) +{\cal O}(a^2)+{\cal O}(g^2a^2) \right], 
	{\hskip 1cm} ({\mathrm{setting}}\ \beta=\frac{6}{g^2}), \nonumber \\
	& =  {\displaystyle a^4 \sum_{x_0} \sum_{\mu,\nu} } & \left[
        \frac{1}{2} {\rm {Tr}} F_{\mu \nu}^2 (x_0) +{\cal O}(a^2)+{\cal O}(g^2a^2) \right],
\label{eq:WilsonF} 
\end{eqnarray}
which reproduces the continuum action to ${\cal O}(a^2)$. We note that the path-ordering is crucial to 
obtaining the non-abelian $F_{\mu \nu} = \dmu \Anu - \dnu \Amu + ig[\Amu,\Anu]$ with errors of order 
${\cal O}(g^2a^2)$.\\

In this investigation we use cooling to eliminate the high-frequency components of the 
gauge fields on the lattice. This reveals topological structures which correspond with 
classical minima of the action. The cooling algorithm works by calculating the local action at 
each lattice link, and updating the link to minimize this action. However the 
discretization errors inherent in the Wilson action cause the cooling algorithm to 
consistently underestimate the action at each lattice link \cite{Perezandco}. As a result 
the link updates do not accurately reflect the true structure of the gauge fields, leading 
to the destabilization of non-trivial topological structures (instantons and anti-instantons).

\subsection{Improving the Action}
The standard Wilson plaquette action, $S_{\rm {Wil}}$, contains deviations from the continuum 
Yang-Mills action of order ${\cal O}(a^2)$ and ${\cal O}(g^2a^2)$. This problem 
may be remedied by improving the action. Tree-level improvement is a simple and effective means 
of eliminating lowest-order 
discretization errors by calculating the action from a combination of, for instance, the 
plaquette, $2a \times a$, and $a \times 2a$ rectangles. Since the plaquette and 
rectangles have different ${\cal O}(a^2)$ errors they may be added in a linear 
combination in such a way that the leading error terms cancel and one is 
left with the term corresponding to the Yang-Mills action 
plus ${\cal O}(a^4)$ and ${\cal O}(g^2a^2)$ errors.\\ 

The form for a tree-level improved action can be easily determined from the quantities
\beqa
\frac{1}{3}Re{\rm {Tr}}W_{\mu \nu}^{(1 \times 1)}(x_0) & = & 
	1 - \frac{a^4 g^2}{6} {\rm {Tr}} F_{\mu \nu}^2 (x_0)
	  - \frac{a^6 g^2}{72}{\rm {Tr}} F_{\mu \nu}(x_0)
        \left(\del_{\mu}^2 + \del_{\nu}^2 \right)F_{\mu \nu}(x_0) \nonumber \\
	& & +{\cal O}(g^2a^8)+{\cal O}(g^4a^6)\, , \label{eq:plaquetteFullExpansion} \\
\frac{1}{3}Re{\rm {Tr}}W_{\mu \nu}^{(2 \times 1)}(x_0) & = &
	1 - \frac{4 a^4 g^2}{6} {\rm {Tr}} F_{\mu \nu}^2 (x_0)
          - \frac{4 a^6 g^2}{72}{\rm {Tr}} F_{\mu \nu}(x_0)
        \left(4\del_{\mu}^2 + \del_{\nu}^2 \right)F_{\mu \nu}(x_0) \nonumber \\
        & & +{\cal O}(g^2a^8)+{\cal O}(g^4a^6)\, , \\
\frac{1}{3}Re{\rm {Tr}}W_{\mu \nu}^{(1 \times 2)}(x_0) & = &
        1 - \frac{4 a^4 g^2}{6} {\rm {Tr}} F_{\mu \nu}^2 (x_0)
          - \frac{4 a^6 g^2}{72}{\rm {Tr}} F_{\mu \nu}(x_0)
        \left(\del_{\mu}^2 + 4\del_{\nu}^2 \right)F_{\mu \nu}(x_0) \nonumber \\
        & & +{\cal O}(g^2a^8)+{\cal O}(g^4a^6)\, .
\eeqa
Henceforth we shall make use of the definition 
\begin{equation}
P_{\mu \nu}^{(m \times n)}(x) \equiv \frac{1}{3}Re{\rm {Tr}}W_{\mu \nu}^{(m \times n)}(x).
\end{equation}
 We can therefore readily see that the ${\cal O}(a^2)$ terms may be eliminated from the action
by making the following construction

\begin{eqnarray*}
S_{\rm {Imp}} & = {\displaystyle \frac{5}{3} \beta \sum_{x} \sum_{\mu<\nu} } & \left[
	   	      \left(1-P_{\mu \nu}^{(1 \times 1)}(x_0)\right) 
	- \frac{1}{20}\left(1-P_{\mu \nu}^{(2 \times 1)}(x_0)\right)
	- \frac{1}{20}\left(1-P_{\mu \nu}^{(1 \times 2)}(x_0)\right)
	\right]\, , \\
	& =  {\displaystyle a^4 \sum_{x} \sum_{\mu,\nu} } & \left[
        \frac{1}{2} {\rm {Tr}} F_{\mu \nu}^2 (x_0)
        +{\cal O}(a^4)+{\cal O}(g^2a^2)\, , \right]{\hskip 1cm} ({\mathrm{setting}}\ \beta=\frac{6}{g^2}), \\
\end{eqnarray*}
which reproduces the continuum action to ${\cal O}(a^4)$ and ${\cal O}(g^2a^2)$.\\

This process eliminates classical error terms of order ${\cal O}(a^2)$ arising from the finite 
lattice spacing. Non-classical ${\cal O}(g^2a^2)$ errors arising from self-couplings of the 
gluon fields may be dealt with to some extent by the process of tadpole improvement \cite{LePage}. 
Due to the different numbers of links in plaquettes, rectangles, and other possible choices 
of Wilson loop, one redefines the value of each link by scaling it by the mean link $u_0$
(to account for the significant tadpole-style self-interactions of the gluon fields 
introduced on the lattice via the link operators). 
In essence $u_0$ is an estimate of the sum of these unwanted tadpole terms, after the leading 
${\cal O}(g)$ terms in the links are factored out \cite{LePage}. Scaling the 
links by the mean link
\beq
U_\mu(x) \rightarrow \frac{U_\mu(x)}{u_0}\, ,
\eeq
improves the accuracy of the expansions above. We use the plaquette measure
\begin{equation}
u_0 = \left<P_{\mu \nu}^{(1 \times 1)}(x) \right>^{\frac{1}{4}}_{x,\mu<\nu},
\label{eq:u_0defn}
\end{equation}
which is updated after every sweep of cooling. When we include tadpole improvement 
factors, to correct for the difference between the numbers of links in the plaquette 
and the rectangular Wilson loops, the improved action takes the long-established form
\cite{LePage}
\beq
S_{\rm {Imp}} = \frac{5}{3} {\displaystyle \beta \sum_{x} \sum_{\mu<\nu} } \left[
	   	      \left(1-P_{\mu \nu}^{(1 \times 1)}(x_0)\right) 
	- \frac{1}{20u_0^2}\left(1-P_{\mu \nu}^{(2 \times 1)}(x_0)\right)
	- \frac{1}{20u_0^2}\left(1-P_{\mu \nu}^{(1 \times 2)}(x_0)\right)
	\right].
\label{eq:2LImpAction}
\eeq

There is no reason why the tree-level improvement scheme needs to stop at this point. In principle 
it can be extended to arbitrary orders. For example, de Forcrand {\em et al}. \cite{DeForc} have used tree-level 
improvement to construct a lattice action which eliminates ${\cal O}(a^4)$ 
errors, by using combinations of up to five Wilson loop operators. These five loops, which we denote as $L_1,...,L_5$, 
have dimensions $(1\times 1)$, $(1\times 2)$, $(2\times 2)$, $(1\times 3)$, and $(3\times 3)$. 
Of course, in the case of rectangular loops $m \neq n$ we average the contribution of the loops 
in each direction. Hence we define
\beqa
L_1 & \equiv & P_{\mu \nu}^{(1 \times 1)}, \nonumber \\
L_2 & \equiv & P_{\mu \nu}^{(2 \times 2)}, \nonumber \\
L_3 & \equiv & \frac{1}{2}\left\{P_{\mu \nu}^{(2 \times 1)} +P_{\mu \nu}^{(1 \times 2)}\right\},
            \nonumber \\
L_4 & \equiv & \frac{1}{2}\left\{P_{\mu \nu}^{(3 \times 1)} +P_{\mu \nu}^{(1 \times 3)}\right\},
            \nonumber \\
L_5 & \equiv & P_{\mu \nu}^{(3 \times 3)}. \nonumber 
\eeqa
The improved action of de Forcrand {\em et al}. can then be written in the form
\beq
S = {\displaystyle \sum_{i=1}^5} \frac{1}{(m^2n^2)} c_i S_i\, ,
\label{eq:DeForcAction}
\eeq
where $S_i$ is the action calculated by inserting the loop $L_i$ in place of $U_{\mu\nu}$ in
the definition of the Wilson action, Eq.~(\ref{eq:WilsonAction}), where $m$ and $n$ are the dimensions of 
the loop $L_i$, and where the $c_i$ are improvement constants which control how much each loop contributes to 
the total improved action. In our work we explicitly include tadpole improvement factors, 
\beq
\frac{1}{u_0^{(2m+2n-4)}},
\eeq
whereas de Forcrand {\em et al}. left these factors as unity in Eq.~(\ref{eq:DeForcAction}). 
Our experience is that tadpole improvement factors are beneficial in the 
early stages of cooling and remain beneficial even as $u_0 \rightarrow 1$. \\

The values of the improvement constants which eliminate order ${\cal O}(a^4)$ 
tree-level error terms are 
\begin{eqnarray}
c_1 & = & (19 - 55 c_5 ) / 9\, , \nonumber \\
c_2 & = & (1 - 64 c_5) / 9\, , \nonumber \\
c_3 & = & (640 c_5 - 64 ) / 45\, , \nonumber \\
c_4 & = & 1/5 - 2 c_5\, , \nonumber 
\end{eqnarray}
where $c_5$ is a free variable. By adjusting $c_5$ we create different ${\cal O}(a^4)$-improved actions, 
which in general will have different ${\cal O}(a^6)$ error terms. This means that we can attempt to minimize 
${\cal O}(a^6)$ errors by tuning $c_5$. If we set $c_5 = 1/10$, then $c_3=c_4=0$ 
creating a ``3-loop'' action, i.e., an action constructed from three of the five loops. Setting $c_5 = 0$ 
we create a ``4-loop'' improved action. To construct a specific 5-loop action, de Forcrand {\em et al}. 
chose to set $c_5 = 1/20$, halfway between the 3-loop and 4-loop values to stabilize single instanton 
sizes under cooling. These so-called 3-loop, 4-loop, and 5-loop improved actions are all free of ${\cal O}(a^4)$ 
errors, and have been shown to preserve (anti-)instantons for many thousands of sweeps. 

\subsection{Topological Charge}
On the lattice, the total topological charge is obtained by summing the charge 
density 
\beq
q(x) = \frac{g^2}{32\pi^2}\epsilon_{\mu\nu\rho\sigma}
{\rm {Tr}}\{F_{\mu\nu}(x)F_{\rho\sigma}(x)\} \, ,
\label{eq:LatQDens}
\eeq
over each lattice site
\beq
Q = {\displaystyle \sum_{x}} q(x)\, ,
\eeq
where $\mu, \nu, \rho, \sigma$ sum over the directions of the 
lattice axes. Further to their construction of the improved action, de Forcrand {\em et al}.  
have used an analogous procedure to define an improved lattice topological charge operator \cite{DeForc}. 
We choose to construct the ${\cal O}(a^4)$-improved topological charge in a different manner.
In the next section we derive an improved version of the field-strength tensor, $F_{\mu\nu}$. 
This improved operator forms the basis of our definition of the improved topological charge.
The derivation of the improved field-strength tensor will serve to  illustrate the general methods used to 
deduce the results already stated (without proof) above for the improved action.

\section{The Lattice Field-Strength Tensor}   
\label{sec:LatFmunu}
Consider the problem of expanding a generic Wilson loop operator. Such an operator 
is defined as the path-ordered exponential of the closed path integral around the loop
\begin{equation}
W_{\mu \nu}^{(m \times n)} = {\cal P}\ e^{ig\oint A.dx}.
\label{eq:LoopOperator}
\end{equation}
As noted earlier, the path ordering is crucial to obtaining the full non-abelian field-strength tensor in 
the action associated with $W_{\mu \nu}^{(m \times n)}$ such that the errors are ${\cal O}(g^2a^2)$. 
Note that mean-field improvement reduces the magnitude of these ${\cal O}(g^2a^2)$ errors 
present early in the cooling process. The ${\cO}(g^2a^2)$ errors are also suppressed via cooling and we do 
not consider them further.\\

We are considering tree-level improvement to remove ${\cO}(a^4)$ and higher errors. 
It is therefore sufficient to work to leading non-vanishing order in $g$, which is equivalent to 
the abelian theory. We hence determine the coefficients in the expansion of any Wilson loop by 
considering the weak coupling limit $g \rightarrow 0$. \\

Rewriting the abelian version of the line integral $\oint A.dx$ in Eq.~(\ref{eq:LoopOperator}) as 
$\oint (A_{\mu}dx_{\mu} + A_{\nu}dx_{\nu})$ we see that we can apply 
the \mbox{two-dimensional} version of Stoke's theorem (i.e., Green's theorem)
\begin{equation}
\oint_{\partial R} J dx + K dy = \int\int_R \left(\frac{\partial K}{\partial x}
		- \frac{\partial J}{\partial y}\right)dx dy\, ,
\end{equation}
where $\partial R$ is the closed contour containing the two-dimensional surface
$R$. Therefore we see that we can readily rewrite our integral around the 
loop as
\begin{equation}
\oint A.dx = \int_{-ma/2}^{ma/2}dx_{\mu}\int_{-na/2}^{na/2}dx_{\nu}
	     \left( \partial_{\mu}A_{\nu}(x_0 + x) 
		- \partial_{\nu}A_{\mu}(x_0 + x) \right).
\label{eq:AfterGreensThm}
\end{equation}
where $x_0$ is the centre of the loop. 
The integrand is identified as the abelian field-strength tensor
\begin{equation}
\oint A.dx = \int_{-ma/2}^{ma/2}dx_{\mu}\int_{-na/2}^{na/2}dx_{\nu}
	     F_{\mu\nu}(x_0 + x)\, .
\label{eq:FmunuIntegrated}
\end{equation}
Hence the expansion of any abelian loop will be in powers of $F_{\mu\nu}$, reproducing 
the $F_{\mu\nu}F^{\mu\nu}$ term of the action at order ${\cO}(g^2)$, 
and containing error terms of relative order ${\cO}(g^2a^2)$ and ${\cO}(a^4)$. \\

Now we Taylor expand the $F_{\mu\nu}$'s around the point $x_0$
\begin{eqnarray}
\oint A.dx & = & \int_{-ma/2}^{ma/2}dx_{\mu}\int_{-na/2}^{na/2}dx_{\nu}
        \left( F_{\mu\nu}(x_0) 
	+ \sum_\alpha x_{\alpha}\del_{\alpha}F_{\mu\nu}(x_0)\frac{}{}
        \right. \nonumber \\
    & & \left. +\frac{1}{2} \sum_{\alpha,\beta} x_{\alpha}x_{\beta}\del_{\alpha}\del_{\beta}F_{\mu\nu}(x_0) 
        +\cdots \right)\, ,  
\label{eq:TaylorExpansion}
\end{eqnarray}
Since our plaquette is a two-dimensional object we sum over the $\mu$ and $\nu$ directions, ($\alpha=\mu$ or 
$\nu$, and similarly for $\beta$).
\begin{eqnarray}
\oint A.dx 
	& = & \int_{-ma/2}^{ma/2}dx_{\mu}\int_{-na/2}^{na/2}dx_{\nu}
            \left( \frac{}{} F_{\mu \nu}(x_0)
	  + \left\{x_{\mu}\del_{\mu} + x_{\nu}\del_{\nu}\right\}
	    F_{\mu \nu}(x_0) \right. \nonumber \\
        & & + \left\{ x_{\mu} x_{\nu}\del_{\mu}\del_{\nu} \right\}
	    F_{\mu \nu}(x_0)
	  + \frac{1}{2}\left\{ x_{\mu}^2 \del_{\mu}^2
          + x_{\nu}^2 \del_{\nu}^2 
          + x_{\mu}^2 x_{\nu} \del_{\mu}^2 \del_{\nu} 
          + x_{\mu} x_{\nu}^2 \del_{\mu} \del_{\nu}^2
            \right\}
	    F_{\mu \nu}(x_0)\nonumber \\
        & & +\frac{1}{6}\left\{x_{\mu}^3 \del_{\mu}^3 + x_{\nu}^3 \del_{\nu}^3 
                        + x_{\mu}^3 x_{\nu} \del_{\mu}^3 \del_{\nu} 
                        + x_{\mu} x_{\nu}^3 \del_{\mu} \del_{\nu}^3 \right\}    
            F_{\mu \nu}(x_0) \nonumber \\
        & & +\frac{1}{24}\left\{x_{\mu}^4 \del_{\mu}^4 + x_{\nu}^4 \del_{\nu}^4 \right\}    
                    F_{\mu \nu}(x_0)
     			\nonumber \\
        & & +\frac{1}{4}\left\{x_{\mu}^2 x_{\nu}^2 \del_{\mu}^2 \del_{\nu}^2 
                + x_{\mu}^2 x_{\nu}^2 \del_{\mu}^2 \del_{\nu}^2 \right\}    
                    F_{\mu \nu}(x_0)
     		    \left. +{\cal O}(x^5)  \right).
\label{eq:expansion}
\end{eqnarray}

The terms with odd powers of $x_\mu$ or $x_\nu$ vanish upon integration, because of 
the symmetric integration limits, about the center of the loop. Let us perform the expansion of 
the standard $(1 \times 1)$ plaquette. 
\begin{eqnarray}
\oint A.dx & = & \int\int_{-a/2}^{a/2}dx_{\mu}dx_{\nu}
        \left(F_{\mu \nu}(x_0) 
	+ \frac{1}{2}\left\{ x_{\mu}^2 \del_{\mu}^2
                          + x_{\nu}^2 \del_{\nu}^2 \right\} F_{\mu \nu}(x_0) \right.
               \nonumber \\
        & & \left. +\frac{1}{24}\left\{x_{\mu}^4 \del_{\mu}^4 + x_{\nu}^4 \del_{\nu}^4 \right\}    
                    F_{\mu \nu}(x_0)
            +\frac{1}{4}\left\{x_{\mu}^2 x_{\nu}^2 \del_{\mu}^2 \del_{\nu}^2 
                + x_{\mu}^2 x_{\nu}^2 \del_{\mu}^2 \del_{\nu}^2 \right\}
               F_{\mu \nu}(x_0)+{\cal O}(x^5) \right)\, ,
               \nonumber \\
	& = & a^2 F_{\mu \nu}(x_0) + \frac{a^4}{24}
	      \left(\del_{\mu}^2 + \del_{\nu}^2 \right)F_{\mu \nu}(x_0) 
               \nonumber \\
        &   & + \frac{a^6}{1920} \left(\del_{\mu}^4 + \del_{\nu}^4 \right)F_{\mu \nu}(x_0)
              + \frac{a^6}{576} \left(\del_{\mu}^2 \del_{\nu}^2 \right)F_{\mu \nu}(x_0)
 	      +{\cal O}(a^8)\, . 
        \label{eq:unsquared}
\end{eqnarray}
This expansion can be determined to arbitrary order by taking the Taylor expansion in 
Eq.~(\ref{eq:TaylorExpansion}) to the desired order. \\

So consider again the form taken by the equation for the expansion of the $(1\times 1)$ loop 
operator in the non-abelian case, which derives from the Schwinger line integral around a closed path
\begin{equation}
W^{(1 \times 1)}_{\mu \nu}  = {\cal P} e^{ig\oint A.dx}  = 
      {\cal P} \left\{1+ig\oint A.dx-\frac{g^2}{2}(\oint A.dx)^2+\cdots \right\}.
\label{eq:W-ExponExpan}
\end{equation}
The standard approach to constructing the action is to take the real part of the trace of this expansion, 
thereby extracting the leading term which is just one, and the third term which is proportional to 
$F_{\mu\nu}^2$. However we are interested here in extracting the term proportional to $F_{\mu\nu}$. We 
can do this by making the following construction, which takes advantage of the hermitian properties of the 
Gell-Mann matrices
\begin{eqnarray}
W^{(1 \times 1)}_{\mu \nu}     & = & 
 	{\cal P} \left\{1+ig\oint A dx -\frac{g^2}{2}(\oint A dx)^2
	 + {\cal O}(g^3)\right\}\, , 
                                                        \nonumber \\ 
W^{(1 \times 1){\dagger}}_{\mu \nu} & = & 
        {\cal P} \left\{1-ig\oint A dx -\frac{g^2}{2}(\oint A dx)^2
         + {\cal O}(g^3)\right\}\, ,
                                                        \nonumber 
\end{eqnarray}
and hence
\begin{eqnarray}
\frac{-i}{2}\left[W^{(1 \times 1)}_{\mu \nu} - 
    	             W^{(1 \times 1){\dagger}} 
              - \frac{1}{3}{\rm {Tr}}(W^{(1 \times 1)}_{\mu \nu} - 
    	                    W^{(1 \times 1){\dagger}}_{\mu\nu})\right]
	 & = & g {\cal P} \oint A dx + {\cal O}(g^3) \nonumber \\
         & = & ga^2 \left[F_{\mu \nu}(x_0) + {\cal O}(a^2) + {\cal O}(g^2a^2)\right]\, . \nonumber \\
         &   &
\label{eq:ExtractIntegral}
\end{eqnarray}
We have subtracted one-third of the trace to enforce the traceless property of the 
Gell-Mann matrices.\\
 
Eq.~(\ref{eq:ExtractIntegral}) demonstrates how we construct the field-strength tensor 
$F_{\mu \nu}$ from the $(1 \times 1)$ plaquette. Since the expansion in Eq.~(\ref{eq:unsquared}) 
corresponds with a given Wilson loop only through the choice of integration limits, this suggests that it is also 
possible to construct an improved field-strength tensor from a suitably chosen linear combination of Wilson loops. 
This improved field-strength tensor can be fed directly into Eq.~(\ref{eq:LatQDens}) 
resulting in a topological charge which is automatically free of 
discretization errors to the same order in $a^2$. Furthermore, since the 
Yang-Mills action is based upon the field-strength tensor (squared), 
it is possible to create a {\em reconstructed action}, $S_R$, by inserting the improved 
field-strength tensor directly into the equation
\beq
S_R = \beta \sum_{x} \sum_{\mu,\nu} \frac{1}{2} {\mathrm Tr} F_{\mu \nu}(x)^2\, .
\eeq
Since the reconstructed action is improved differently
to the standard improved action operator, Eq.~(\ref{eq:DeForcAction}), 
and constructed from clover terms, the value of the action calculated with the 
reconstructed action operator can be compared with the value calculated with the standard 
improved action operator at each cooling sweep as a mechanism to explore the size of 
the remaining discretization errors.\\

\section{Improving The Field-Strength}
\label{sec:ImpFmunu}
\begin{figure}[t]
\begin{center}
\leavevmode
\epsfig{figure=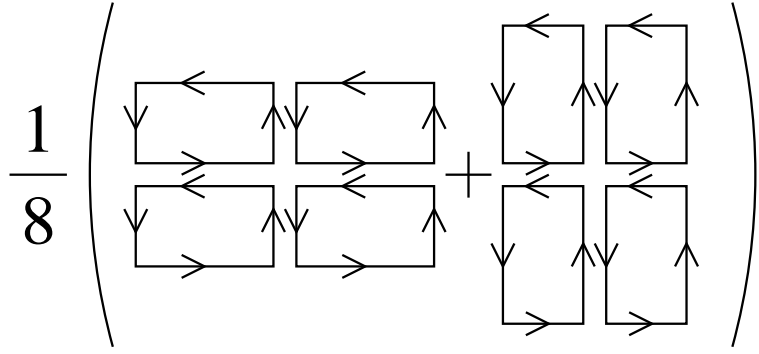,angle=0,width=10cm}
\end{center}
\caption{The four $m \times n$ loops and four $n \times m$ loops used to construct the 
clover term from which we construct the lattice topological charge operator. \newline}
\label{fig:cloverloops}
\end{figure}

We now wish to construct sums of clover terms designed to remove ${\cO}(a^2)$ and ${\cO}(a^4)$
errors relative to the leading term, $a^2 F_{\mu \nu}$. Let us begin by defining
\begin{eqnarray}
{\cal A} & = & ga^2 F_{\mu \nu}\, , \nonumber \\
{\cal B} & = & ga^4 \left(\del_{\mu}^2 + \del_{\nu}^2 \right)F_{\mu \nu}\, , \nonumber \\ 
{\cal C} & = & ga^6 \left(\del_{\mu}^4 + \del_{\nu}^4 \right)F_{\mu \nu}\, , \nonumber \\
{\cal D} & = & ga^6 \left(\del_{\mu}^2 \del_{\nu}^2 \right)F_{\mu \nu}\, .  
\end{eqnarray}
Let us now denote $C^{(m,n)}$ as the combination of 
Wilson loop terms extracted from the left-hand-side of Eq.~(\ref{eq:ExtractIntegral}) 
corresponding with the loops used to construct a clover term as depicted in  
Fig.~\ref{fig:cloverloops}. 
Constructing the 
four clover terms and symmetrizing in $n \leftrightarrow m$ we have 
\begin{eqnarray}
C^{(m,n)} & = {\displaystyle \frac{1}{8}} & \left\{
                          \int_{0}^{ma}dx_{\mu}\int_{0}^{na}dx_{\nu}F_{\mu \nu} 
                        + \int_{-ma}^{0}dx_{\mu}\int_{0}^{na}dx_{\nu}F_{\mu \nu} \right. \nonumber \\
       & &                 + \int_{-ma}^{0}dx_{\mu}\int_{-na}^{0}dx_{\nu}F_{\mu \nu}
                           + \int_{0}^{ma}dx_{\mu}\int_{-na}^{0}dx_{\nu}F_{\mu \nu}\nonumber \\
       & &                 + \int_{0}^{na}dx_{\mu}\int_{0}^{ma}dx_{\nu}F_{\mu \nu}
                           + \int_{-na}^{0}dx_{\mu}\int_{0}^{ma}dx_{\nu}F_{\mu \nu}\nonumber \\
       & &       \left.    + \int_{-na}^{0}dx_{\mu}\int_{-ma}^{0}dx_{\nu}F_{\mu \nu}
                           + \int_{0}^{na}dx_{\mu}\int_{-ma}^{0}dx_{\nu}F_{\mu \nu}\right\}, \nonumber \\
       & = {\displaystyle \frac{1}{8}} & \left\{ \int_{-ma}^{ma}dx_{\mu}\int_{-na}^{na}dx_{\nu}F_{\mu \nu}
                           + \int_{-na}^{na}dx_{\mu}\int_{-ma}^{ma}dx_{\nu}F_{\mu \nu}\right\}.
\end{eqnarray}
Expanding the loop operators as in Eq.~(\ref{eq:expansion}) we find that 
\begin{eqnarray}
C^{(1,1)} & = & 
  {\cal A}  +  \frac{1}{6}{\cal B}  +  \frac{1}{120}{\cal C} 
            +  \frac{1}{36}{\cal D}\, ,  
               \nonumber \\ 
          &   & \nonumber \\ 
C^{(2,2)} & = & 
  4{\cal A} +  \frac{8}{3}{\cal B}  +  \frac{8}{15}{\cal C} 
            +  \frac{16}{9}{\cal D}\, ,  
               \nonumber \\
          &   & \nonumber \\
C^{(1,2)} & = & 
  2{\cal A} +  \frac{5}{6}{\cal B}  +  \frac{17}{120}{\cal C} 
            +  \frac{2}{9}{\cal D}\, ,  
               \nonumber \\
          &   & \nonumber \\
C^{(1,3)} & = & 
  3{\cal A} +  \frac{5}{2}{\cal B}  +  \frac{41}{40}{\cal C} 
            +  \frac{3}{4}{\cal D}\, ,  
               \nonumber \\
          &   & \nonumber \\
C^{(3,3)} & = & 
  9{\cal A} +  \frac{27}{2}{\cal B}  +  \frac{243}{40}{\cal C} 
            +  \frac{81}{4}{\cal D}\, .  
                \label{eq:go-ban}  
\end{eqnarray}
Notice that the ${\cal O}(a^6)$ terms in each $F_{\mu \nu}$ will be ``promoted''
to ${\cal O}(a^8)$ (and higher) by taking the product $F_{\mu \nu}F_{\rho \sigma}$, while 
the ${\cal O}(a^8)$ terms will be promoted to ${\cal O}(a^{10})$ (and higher). 
Consequently, since we wish to eliminate ${\cal O}(a^8)$ errors from the product, we need only 
expand out to (and eliminate) order ${\cal O}(a^6)$ corrections in $F_{\mu \nu}$. We therefore 
have five equations and four unknowns.\\

The improved field-strength may be written
\begin{equation}
F^{\mathrm {Imp}}_{\mu\nu} =
		k_1 C^{(1,1)}_{\mu \nu} + 
		k_2 C^{(2,2)}_{\mu \nu} +
		k_3 C^{(1,2)}_{\mu \nu} + 
		k_4 C^{(1,3)}_{\mu \nu} + 
		k_5 C^{(3,3)}_{\mu \nu}, 
\end{equation}
where the $k_i$, the ``weightings'' of each loop, are defined by
\begin{equation}
\begin{array}{ccccccccccc}
k_1 {\cal A}^{(1,1)}& +& k_2 {\cal A}^{(2,2)}& +& k_3 {\cal A}^{(1,2)}& + 
                       & k_4 {\cal A}^{(1,3)}& +& k_5 {\cal A}^{(3,3)}& = & 1\, , \label{eq:alpha} \\
k_1 {\cal B}^{(1,1)}& +& k_2 {\cal B}^{(2,2)}& +& k_3 {\cal B}^{(1,2)}& + 
                       & k_4 {\cal B}^{(1,3)}& +& k_5 {\cal B}^{(3,3)}& = & 0\, , \\
k_1 {\cal C}^{(1,1)}& +& k_2 {\cal C}^{(2,2)}& +& k_3 {\cal C}^{(1,2)}& + 
                       & k_4 {\cal C}^{(1,3)}& +& k_5 {\cal C}^{(3,3)}& = & 0\, , \\
k_1 {\cal D}^{(1,1)}& +& k_2 {\cal D}^{(2,2)}& +& k_3 {\cal D}^{(1,2)}& + 
                       & k_4 {\cal D}^{(1,3)}& +& k_5 {\cal D}^{(3,3)}& = & 0\, , \label{eq:omega}
\end{array}
\end{equation}
such that the weighted coefficient of $g a^2 F_{\mu \nu}$ is 1 and the weighted coefficients 
of the ${\cal O}(a^2)$ and ${\cal O}(a^4)$ terms vanish in the sum of loops.
Here ${\cal A}^{(m,n)}$ represents the coefficient of the ${\cal A}$ term from the 
expansion of $C_{\mu \nu}^{(m,n)}$ and so forth, with the values of the coefficients 
taken from Eqs.~(\ref{eq:go-ban}). To find the values of the 
improvement constants we use these coefficients to construct an equivalent matrix 
equation. Eq.~(\ref{eq:alpha}) takes the form
\begin{equation}
\left[ \begin{array}{cccccc}
       1      &       4        &       2        &        3       &        9 \\ 
              &                &                &                &            \\ 
\frac{1}{6}   &\frac{8}{3}     &\frac{5}{6}     &\frac{5}{2}     &\frac{27}{2} \\ 
              &                &                &                &            \\ 
\frac{1}{120} &\frac{8}{15}    &\frac{17}{120}  &\frac{41}{40}   &\frac{243}{40} \\ 
              &                &                &                &            \\ 
\frac{1}{36}  &\frac{16}{9}    &\frac{2}{9}     &\frac{3}{4}     &\frac{81}{4} 
\end{array}
\right] 
\left[ \begin{array}{c}
k_1 \\ k_2 \\ k_3 \\ k_4 \\ k_5 \end{array}
\right] = 
\left[ \begin{array}{c}
1 \\
0 \\
0 \\
0 \end{array}
\right]\, .
\end{equation}
Using Gauss-Jordan elimination, the improvement coefficients are  
\begin{eqnarray}
k_1 & = & 19/9 - 55 k_5\, , \nonumber \\
k_2 & = & 1/36 - 16 k_5\, , \nonumber \\
k_3 & = & 64 k_5 - 32/45\, , \nonumber \\
k_4 & = & 1/15 - 6 k_5\, , \nonumber 
\end{eqnarray}
where the coefficient of the $3\times 3$ loop (in this case $k_5$) is a tunable 
free parameter. \\

We can see that if we set $k_5 = 1/90$, then $k_3=k_4=0$, eliminating 
the contribution from the $C^{(1,2)}$ and $C^{(1,3)}$ loops and creating a 3-loop 
${\cO}(a^4)$-improved field-strength tensor. We may create a 4-loop improved 
field-strength tensor in three different ways, by setting $k_5=0$, $k_5=19/495$, or $k_5=1/576$, 
where the $3\times 3$, $1\times 1$ or $2\times 2$ loops are eliminated respectively. 
We focus on a 4-loop improved tensor with $k_5=0$ throughout this investigation, 
analogous to the 4-loop action of de Forcrand {\em et al.}\\

It must, of course, be noted that the 3-loop and 4-loop action and field-strength operators are 
simply special cases of the 5-loop operators, corresponding with particular choices of 
the parameters $c_5$ and $k_5$ respectively for the action and field-strength tensor. As noted above, 
we have already chosen to investigate a 5-loop action by using de Forcrand {\em et al.}'s choice of 
$c_5=1/20$, halfway between the 3-loop and 4-loop values of $c_5$. In an analogous manner we select
the value $k_5=1/180$ (midway between the 3-loop and 4-loop values) for the 5-loop improved 
field-strength tensor. 

\section{Comparison of Improvement Schemes}
\label{sec:Analysis}
The configurations used in this work are constructed using the Cabbibo-Marinari \cite{CabMar} 
pseudo-heatbath algorithm with three diagonal $SU(2)$ subgroups looped over twice. We 
thermalize for 5000 sweeps with an ${\cal O}(a^2)$-improved action with tadpole improvement 
from a cold start (all links set to the identity) and select configurations every 500 sweeps 
thereafter. Cooling proceeds via updates of the three diagonal $SU(2)$ subgroups, looped over 
twice \cite{Adel1}. Configurations are numbered consecutively in the order that they were saved 
during the process of thermalization. Hence configuration 1 was saved after 5000 thermalization 
sweeps from a cold start, configuration 2 was saved 500 sweeps after configuration 1, 
configuration 3 was saved 500 sweeps after configuration 2, and so on. Our results are generated 
on a $12^3 \times 24$ (untwisted) periodic lattice at $\beta = 4.60$, with a lattice spacing of 
$a=0.125$ fm. \\

As a thermalized configuration is cooled over the course of many hundreds of sweeps, the action 
of the configuration monotonically decreases. This occurs because the cooling algorithm smooths 
out short-range fluctuations in the field. As these ultraviolet components of the field are 
suppressed, the underlying medium and long-distance structure of the field is revealed. When there are 
no regions of positive and negative topological charge density in the process of annihilating, 
the configuration can be thought of as becoming a dilute instanton gas. If cooling proceeds for long enough, 
the only non-trivial parts of the field which will remain are instantons or anti-instantons. \\

The total action and topological charge of a configuration satisfy the 
condition \cite{ChengLi}
\beq
S \geq |Q|S_0 = |Q|\frac{8\pi^2}{g^2}\, .
\label{eq:SQinequality}
\eeq
Since all continuum single (anti-)instantons are non-trivial minima of 
the local action with $|Q|=1$, they saturate the equality in Eq.~(\ref{eq:SQinequality}) and 
therefore have an associated action of $S_0 = 8\pi^2 / g^2$. 
For well-cooled configurations it follows that we may estimate the number of instantons 
on the lattice by dividing the total 
action by the quantity $S_0=8\pi^2 / g^2$. We shall see that one of the principal 
criteria by which we may judge the value 
of an improvement scheme is how closely the values of $S/S_0$ and the topological charge  
approach integer values as we continue cooling and approach self-duality.\\

In a sufficiently dilute 
instanton gas the only contributions to the action and topological charge of the field will 
be that due to the (anti-)instantons. Hence if there are $n_I$ instantons and $n_A$ anti-instantons 
well-separated on a large volume, the total action and topological charge should be integers, 
and satisfy the conditions $S/S_0=n_I+n_A$ and $Q=n_I-n_A$. As cooling proceeds, 
instanton-anti-instanton pairs will annihilate until $S/S_0 = |Q|$ at which point the 
configuration is self-dual. What we expect to see is that as we begin cooling, 
$Q$ should relatively quickly become integer and remain stable thereafter. As we continue to 
cool towards self-duality we should then see $S/S_0$ monotonically decreasing to the point 
where $S/S_0 \rightarrow |Q|$ from above.\\ 

Our lattice volume is not large enough that we observe configurations, stable under cooling, 
with both instantons and anti-instantons simultaneously, i.e., on 
\begin{figure}[t]
\begin{center}
\epsfig{figure=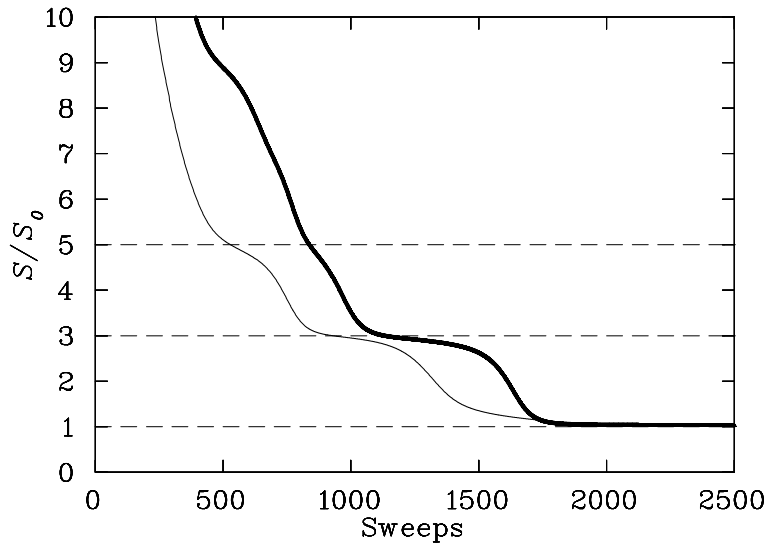,angle=90,width=10cm}
\end{center}
\caption{Action values of configuration 1 (thin line) and configuration 11 (bold line), cooled 
with the 3-loop improved action. Both configurations have a total topological charge of 1. 
It can be clearly seen that the action of each configuration repeatedly plateaus briefly near 
integer values (dashed lines) before dropping by two units. This is characteristic of 
instanton-anti-instanton annihilation events. In the case of configuration 11, plateauing begins 
around sweep 500 ($S/S_0 \approx 9$). \newline}
\label{fig:cfg01and11plateaus}
\end{figure}
our lattice they annihilate as cooling proceeds. In Fig.~\ref{fig:cfg01and11plateaus} we present the action for 
two configurations each having a topological charge (stabilized after preliminary cooling) of $|Q|=1$. 
As cooling proceeds one observes that $S/S_0$ repeatedly decreases by an increment of two. During this 
period the topological charge remains stable. This indicates that the action is dropping due to 
instanton-anti-instanton annihilation events. Notice that the plateaus persist for longer as 
cooling continues. It is known that there can be no fully self-dual $|Q|=1$ configurations on the torus 
\cite{Taubes,Schenk}. The fate of $|Q|=1$ configurations under continued cooling will be discussed elsewhere 
\cite{UsNahm}. We have observed that configurations which stabilize at larger topological charges do not 
exhibit such pronounced plateauing. It is reasonable to deduce that as the lattice becomes more sparsely 
populated, the mean period between instanton-anti-instanton annihilation events becomes longer, as it becomes 
possible to have well-separated, weakly-interacting instantons and anti-instantons on the lattice. On a sufficiently 
large lattice, for sufficiently small total $|Q|$ it should then be possible to arbitrarily closely approach a true 
dilute instanton gas with both instantons and anti-instantons, where the configuration is locally (but not globally) 
self-dual.\\

\begin{figure}[t]
\begin{center}
\epsfig{figure=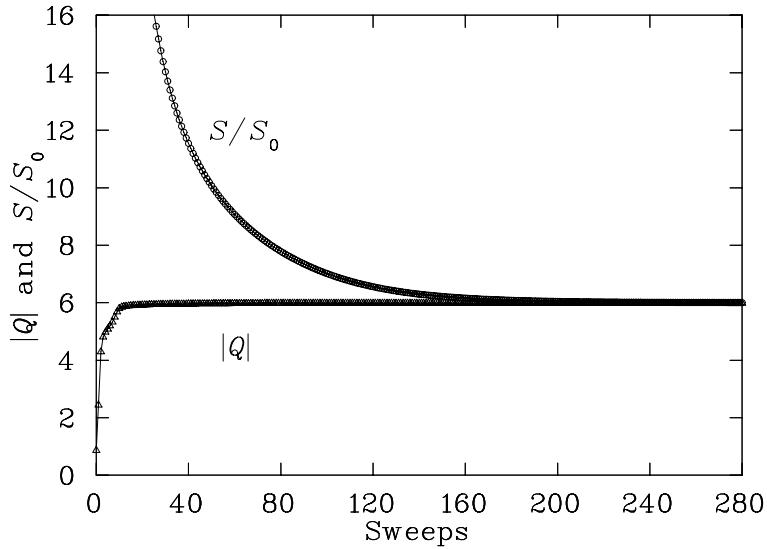,angle=90,width=10cm}
\end{center}
\caption{An example plot of how improved cooling stabilizes the action (circles) and 
topological charge (triangles) at consistent values. The action is 
rescaled by dividing by $S_0$, the action of a single instanton. Cooling is performed 
with an $S(3)$ action, and the topological charge is $Q(3)$. [The notation used is that $S(n)$, 
$S_R(n)$, and $Q(n)$ are the $n$-loop forms of the improved action, our reconstructed action,
and our improved topological charge operator respectively.] 
\newline}
\label{fig:cooling}
\end{figure}

On the lattice, discretization errors are expected to lead to non-integer values of $S/S_0$ and 
topological charge for a highly-cooled configuration. The better the improvement, the closer these 
values will come to integers. 
Fig.~\ref{fig:cooling} shows how $S/S_0$ and the topological charge 
approach the same value, 
indicating the approach to a self-dual configuration. They remain stable for several 
hundred sweeps of improved cooling. We now wish to determine which cooling scheme 
(3-loop, 4-loop, or 5-loop) and which field-strength improvement scheme are best.\\ 

Throughout the discussion of our results we shall use the following notation: the cooling 
action will be denoted by $S$, the reconstructed action by $S_R$ and our highly-improved 
topological charge by $Q$. In some cases the type of improvement scheme used will be denoted by a number in 
parentheses. Hence the 3-loop improved cooling action is written as $S(3)$, our topological 
charge calculated from a 4-loop improved field-strength tensor is written as $Q(4)$, our 
5-loop reconstructed action is written as $S_R(5)$, and so on.

\subsection{Improved Cooling}
\label{subsec:ImpCool}
We shall commence the assessment of which improved action has the smallest discretization 
errors by demonstrating explicitly the discretization errors in the standard Wilson (plaquette) action. 
\begin{figure}[t]
\begin{center}
\epsfig{figure=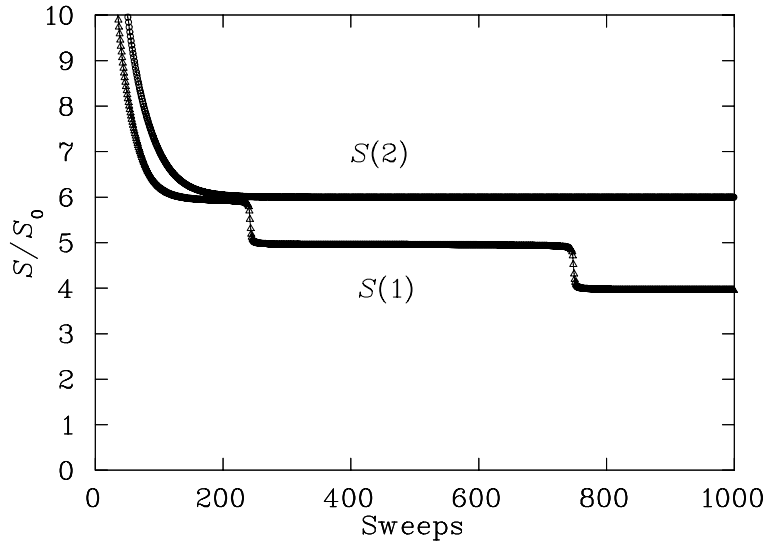,angle=90,width=10cm}
\caption{Action of configuration 89 cooled with $S(1)$ (triangles) and $S(2)$
(circles) cooling schemes after 1000 sweeps. [See Fig.~(\ref{fig:cooling}) caption for notation.]}
\label{fig:ActionCooling1Land2L-cfg89}
\end{center}
\end{figure}
Fig.~\ref{fig:ActionCooling1Land2L-cfg89} shows the action against 
sweep number for a configuration cooled with the Wilson action (triangles), denoted as $S(1)$, 
and the $S(2)$ action (circles) from Eq.~(\ref{eq:2LImpAction}). It can be clearly seen that the 
Wilson action drops to a temporary plateau, but eventually destabilizes and drops by approximately 
one unit around sweep number 250, and again around sweep number 750. It should also be noted 
that when the numerical data are examined closely these plateaus are found to occur somewhat below 
integer values. However the $S(2)$ cooling scheme plateaus at a value much closer to integer 
(in this case $6$) and remains at this value without destabilizing after 1000 cooling sweeps. \\

This result (and other similar results on other 
configurations) are an indication that the ordinary Wilson plaquette action significantly underestimates 
the true action of the configuration. Because the large discretization errors in the action prevent 
the cooling algorithm from updating the links in a manner which accurately reflects the classical 
structure of the fields, plaquette cooling destroys important topological 
structures. It is true that the action may be expected to drop (while the topological charge  
remains constant) due to instanton-anti-instanton annihilation during the cooling process. 
However, since the action drops by an increment of one unit (not two) it is clear that we are 
not observing a decrease in the action as a result of instanton-anti-instanton annihilation, but 
rather the destabilization and destruction of a single instanton or anti-instanton.\\

While the $S(2)$ results are promising, it has already been shown analytically that $S(2)$ does not 
stabilize instantons on the lattice \cite{Perezandco}. Removal of ${\cal O}(a^4)$ errors and long-term 
stabilization of instantons requires the consideration of additional loops.\\

We now shift our attention to comparisons between the various improved actions. 
Figs.~\ref{fig:cfg89action-compare-zin} and \ref{fig:cfg32c2-5} show the value of the action 
attained by two different initial hot configurations as they are cooled with various improved 
actions.
We can see that in both cases the $S(2)$ action plateaus well below the relevant integer 
value, due to large ${\cal O}(a^4)$ discretization errors with negative sign. We furthermore 
notice that in the case of Fig.~\ref{fig:cfg89action-compare-zin}, the $S(4)$ action 
drops below the integer plateau at $S/S_0=6$. In fact, while not shown here it continues falling 
to a plateau at $S/S_0=5$, indicating that it cools away an instanton-like structure which 
the other actions do not. When we examine Fig.~\ref{fig:cfg32c2-5} we see that the 
discretization errors in the $S(4)$ action are relatively large and have negative sign. 
This suggests that the $S(4)$ action consistently underestimates 
the action of the configuration to which it is applied. In the case of 
Fig.~\ref{fig:cfg89action-compare-zin} this underestimation was severe enough to 
completely destroy an instanton during the cooling process.
This in itself is quite a remarkable result, since it demonstrates that ${\cal O}(a^6)$ 
discretization errors can be large enough to destabilize instantons in some cases. We shall 
henceforth discard $S(4)$ cooling on the basis that it has large ${\cO}(a^6)$ discretization 
errors with negative sign.\\

\subsection{Improved Topological Charge and Reconstructed Action}
\label{subsec:ImpTopQ}
\begin{figure}[p]
\begin{center}
\epsfig{figure=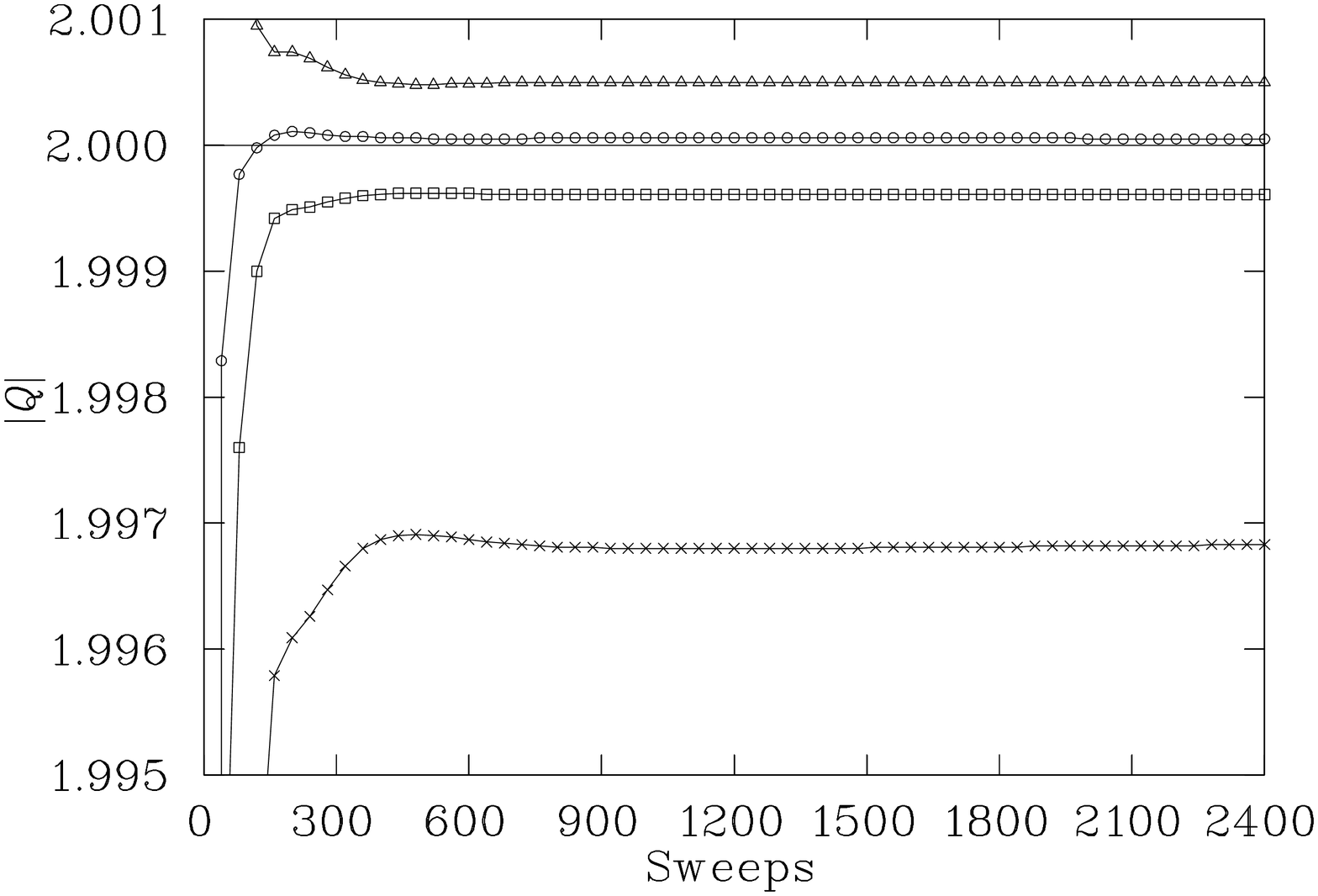,angle=90,width=12cm}
\end{center}
\begin{center}
\epsfig{figure=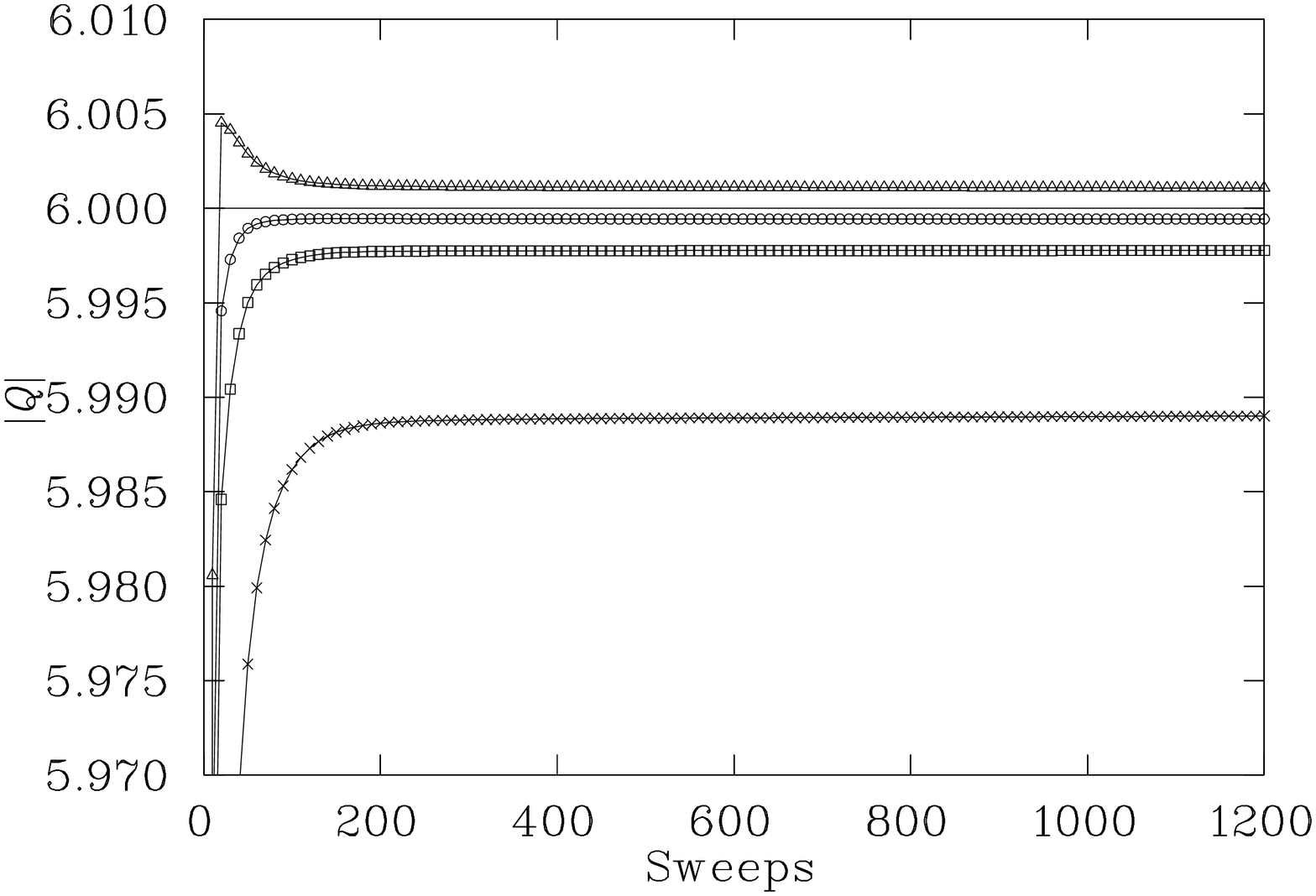,angle=90,width=12cm}
\vspace{5mm}
\caption{Topological Charge of configuration 27 (top) and configuration 89 (bottom) cooled 
exclusively with the $S(3)$ cooling scheme, for 2400 sweeps and 1200 sweeps respectively. In 
descending order the curves are $Q(3)$, $Q(5)$, $Q(4)$, and $Q(2)$.
The $Q(5)$ (circles) operator falls directly between the $Q(3)$ (triangles) and $Q(4)$ (squares) 
topological charge operators, at almost perfect integer values. [See Fig.~(\ref{fig:cooling}) 
caption for notation.]}
\label{fig:cfg27and89TQcomparison2345}
\end{center}
\end{figure}

In order to determine which improvement scheme for $F_{\mu\nu}$ produces the most continuum-like 
results we will assess which topological charge operator produces results closest to integer 
values. The top diagram of Fig.~\ref{fig:cfg27and89TQcomparison2345} shows the development 
of the topological charge of a hot configuration over the course of 2400 cooling 
\begin{figure}[p]
\begin{center}
\epsfig{figure=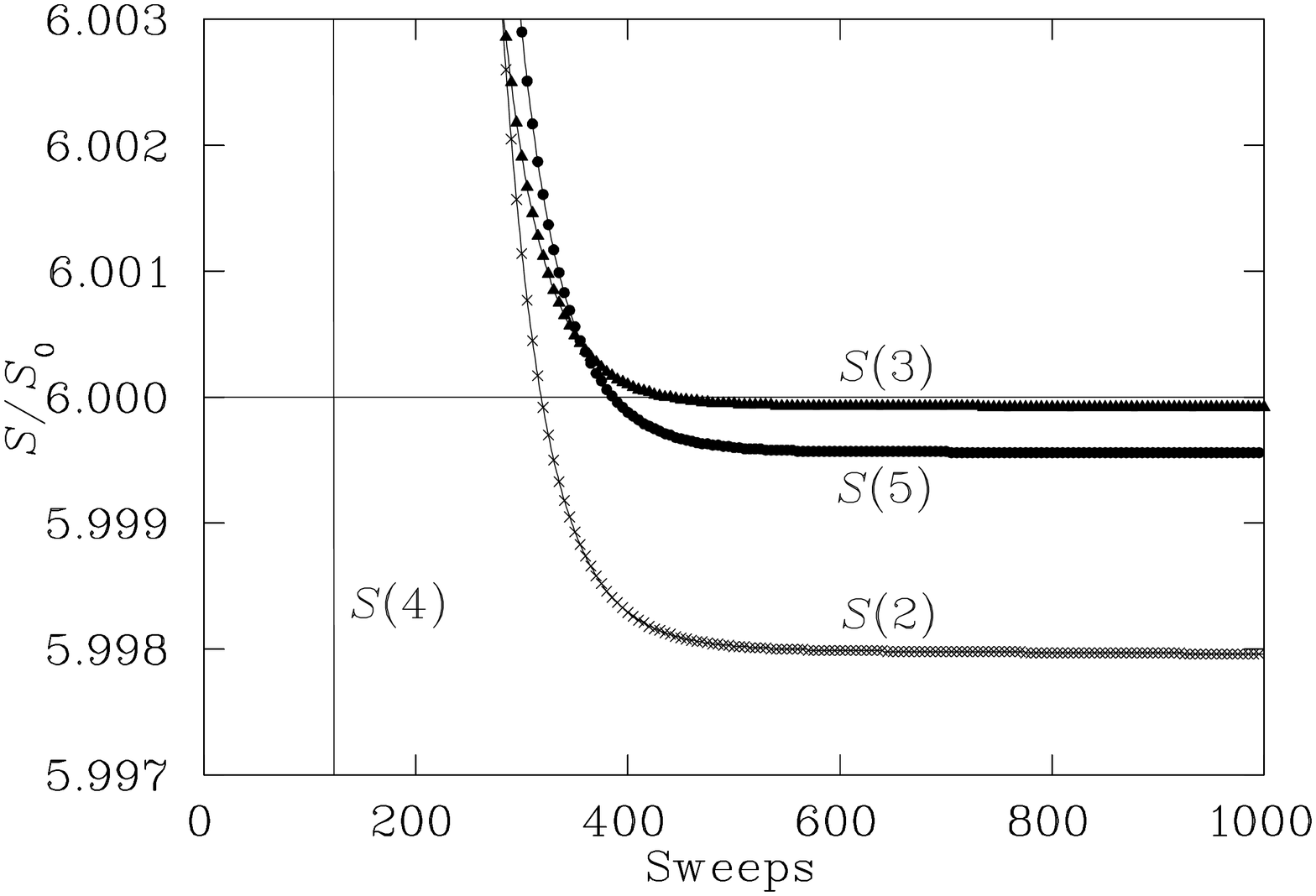,angle=90,width=12cm}
\end{center}
\caption{Action values over the first 1000 cooling sweeps of configuration 89 cooled with the 
$S(2)$ (crosses), $S(3)$ (triangles), $S(4)$ (vertical line) and $S(5)$ (circle) actions. Note 
that $S(4)$ drops to a value near 5.00, while the other actions plateau near 6.00.
[See Fig.~(\ref{fig:cooling}) caption for notation.]
\newline}
\label{fig:cfg89action-compare-zin}
\begin{center}
\epsfig{figure=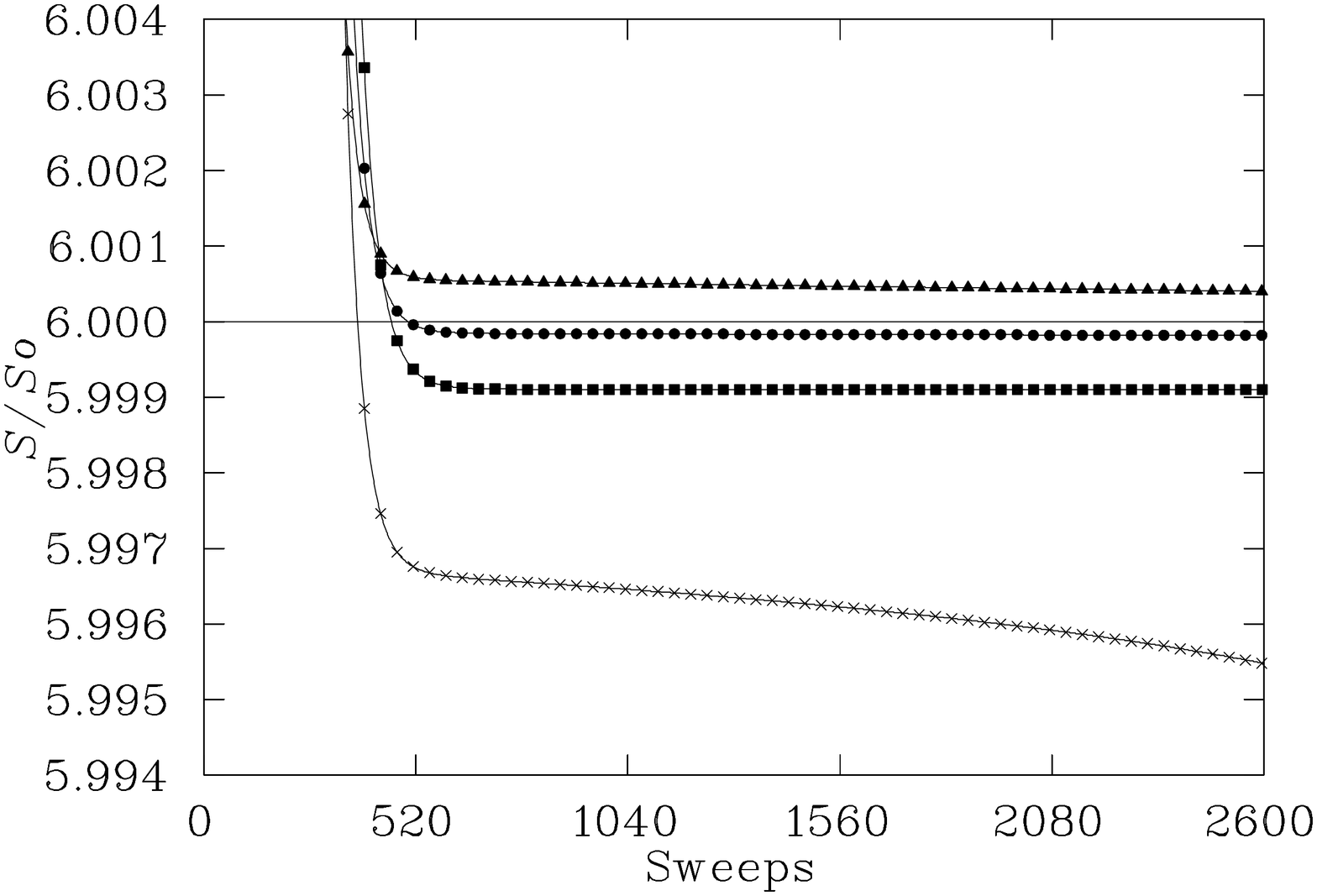,angle=90,width=12cm}
\end{center}
\caption{Action values over the first 2600 cooling sweeps of configuration 32 cooled with 
(from top to bottom on the right-hand-side of the figure) $S(3)$ (triangles), $S(5)$ 
(circles), $S(4)$ (squares), and $S(2)$ (crosses). 
[See Fig.~(\ref{fig:cooling}) caption for notation.]\newline}
\label{fig:cfg32c2-5}
\end{figure}
sweeps with the $S(3)$
improved cooling action. The $Q(1)$ topological charge is too far from an integer 
value to be seen on the scale we have chosen for this diagram, but we can clearly see that 
the $Q(3)$ and $Q(4)$ operators produce significantly better results than the $Q(2)$ 
operator. At the bottom we see equivalent results for configuration 89 (used in 
Fig.~\ref{fig:cfg89action-compare-zin}), over 1200 sweeps. This time we see that the 
$Q(3)$ operator gives marginally more integer-like results than 
those observed with the $Q(4)$ operator. But again the $Q(1)$ and $Q(2)$ operators are 
substantially more inaccurate, as we would expect from the order of improvement used to construct 
$Q(3)$ and $Q(4)$. Our $Q(5)$ operator, constructed with a value of $k_5$ midway between the 
3-loop and 4-loop values, produces near-perfect results for both parts of 
Fig.~\ref{fig:cfg27and89TQcomparison2345}, sitting between the $Q(3)$ and $Q(4)$ 
values. Further fine tuning of $k_5$ appears unnecessary as $Q(5)$ lies above and below the 
integer values in the top and bottom plots of Fig.~\ref{fig:cfg27and89TQcomparison2345} 
respectively. \\

Since the improved field-strength tensor plays 
no role in affecting the structure of the field configurations as they cool, we consider an 
ensemble of configurations that have already been cooled to self-duality (for several hundred 
sweeps) and assess their topological charge (and reconstructed action) with each different improved 
$F_{\mu\nu}$. Table~\ref{tbl:compareQ} shows these results for a number of configurations, all $S(3)$ cooled. 
We can see that in three of the four cases the 5-loop improved topological charge, $Q(5)$, and reconstructed action,
$S_R(5)$, produce results that are both closest to integer values, and in closest agreement to the value 
of the cooling action. 
In the case of configuration 56, $S_R(3)$ and $Q(3)$ produce the results closest to integer 
values.  \\

A further comparison, between results for $S(3)$ and $S(5)$ cooling, is presented in 
Table~\ref{tbl:compareReconS}. In this case we wish to determine whether the choice of cooling 
action affects the dependability of the reconstructed action. We have not chosen to consider 
the $S(4)$ cooling as it has already been deemed unreliable. We see that in the later stages of 
cooling, $S_R(5)$ gives values closer to the cooling action used in each case (whether the 
cooling action is 3-loop or 5-loop), which suggests that it may be a more accurate probe of the 
structure of the fields produced by the cooling algorithm than either the 3-loop or 4-loop 
reconstructed action operators. For these reasons it appears that the 5-loop improved field-strength 
tensor with $k_5=1/180$ is the most dependable of the choices we have examined.\\ 
  
\section{Discussion}
\label{sec:Discussions}
We have now considered the relative merits of the various improvement schemes for the  
cooling action. We have also considered the relative merits of the various improvement 
schemes for the reconstructed action and the topological charge operators (which of course 
are indicative of the accuracy of the improved field-strength tensor). 
It appears that the 5-loop improved field-strength is most accurate,
while the 3-loop field-strength is substantially cheaper (in a computational sense) and still 
\begin{table}[t]
\caption{Reconstructed action (rescaled by $S_0$) and Q values from various configurations 
with $S(3)$ cooling. The columns are the configuration 
number, the sweep number at which the calculation was performed, the value of the cooling action, 
the values of $S_R(2)$ to $S_R(5)$, and the values of $Q(2)$ to $Q(5)$. [See Fig.~(\ref{fig:cooling}) 
caption for notation.]}
\begin{center}
\begin{ruledtabular}
\begin{tabular}{ccccccc} 
Config.& Sweep & $S(3)$ & $S_R(2)$ & $S_R(3)$ & $S_R(4)$ & $S_R(5)$ \\ \hline
  89   & 1071  & 5.99992    & 5.98908  & 6.00110  & 5.99777  & 5.99943  \\
  27   & 1400  & 2.00013    & 1.99687  & 2.00055  & 1.99965  & 2.00010  \\
  90   & 1046  & 1.99999    & 1.99849  & 2.00012  & 1.99978  & 1.99995  \\
  56   & 2446  & 1.99996    & 1.99878  & 2.00002  & 1.99979  & 1.99990  \\
\end{tabular}
\end{ruledtabular}
\vspace{3mm}
\begin{ruledtabular}
\begin{tabular}{ccccccc}
Config.& Sweep & $S(3)$ & $Q(2)$ & $Q(3)$ & $Q(4)$ & $Q(5)$  \\ \hline
  89   & 1071  & 5.99992    & 5.98899& 6.00109& 5.99777& 5.99943 \\
  27   & 1400  & 2.00013    & 1.99680& 2.00050& 1.99961& 2.00006 \\
  90   & 1046  & 1.99999    & 1.99848& 2.00011& 1.99977& 1.99994 \\
  56   & 2446  & 1.99996    & 1.99878& 2.00002& 1.99979& 1.99990 \\
\end{tabular}
\end{ruledtabular}
\label{tbl:compareQ}
\end{center}
\end{table}
produces excellent results.\\

Assessment of improvement in the cooling action is more difficult. It has been stated by 
de Forcrand {\em et al}. \cite{DeForc} that $S(5)$ cooling is most dependable and produces 
the best stabilization of instantons. We have cooled configurations  
with a topological charge of $|Q|=2$ and have seen that the instantons 
rapidly settle on mutually consistent sizes, and furthermore each 
instanton changes size by a factor of no more than 1.1 over the course of several thousand 
sweeps. This work will be reported in detail elsewhere \cite{UsNahm}. However de Forcrand {\em et al}.'s work 
establishes a precedent which must be 
recognized, so let us mention some differences between their work and ours.
Firstly, our work is performed on configurations which contain several instantons or 
anti-instantons while de Forcrand {\em et al}.'s work is performed on configurations which consist 
of a single instanton, obtained via $S(1)$ cooling. Furthermore our work is performed on a 4-toroidal 
mesh of lattice points, with untwisted boundary conditions while de Forcrand {\em et al}.'s work is 
performed with twisted boundary conditions. The twist in the boundary conditions serves to stabilize 
the single instanton present, since self-dual $|Q|=1$ configurations are not permitted on an untwisted 
torus \cite{Taubes,Schenk,BraamBaal}.\\

As demonstrated in Table~\ref{tbl:compareReconS}, for our (periodic) lattice the $S(3)$ cooling action 
produces results as good as $S(5)$.
\begin{table}[t]
\caption{Reconstructed action and cooling action values (all rescaled by $S_0$) from 
various configurations cooled with $S(3)$ (columns 3 to 5) and $S(5)$ (columns 6 to 8). 
[See Fig.~(\ref{fig:cooling}) caption for notation.]
After no more than 170 cooling sweeps, the mean-field improvement factor $u_0$ for configurations 89 and 32 
have settled on a value of 0.99992, and remain completely stable thereafter, while configuration 39 settles on 
a value of $u_0=0.99996$. The resultant weighting of the $3 \times 3$ loop to the $1 \times 1$ loop is 
$u_0^8 \approx 0.9995$.}
\begin{center}
\begin{ruledtabular}
\begin{tabular}{cc|ccc|ccc} 
Config.& Sweep & $S(3)$    & $S_R(3)$   & $S_R(5)$  & $S(5)$     & $S_R(3)$  & $S_R(5)$  \\ \hline
  89   &  200  & 6.03297{} & 6.03417{}  & 6.03243{} & 6.05030{}  & 6.05190{} & 6.05012{} \\
       &  600  & 6.00010{} & 6.00130{}  & 5.99961{} & 5.99957{}  & 6.00116{} & 5.99945{} \\
       &  1000 & 5.99992{} & 6.00110{}  & 5.99944{} & 5.99956{}  & 6.00115{} & 5.99945{} \\ \hline
  39   &  220  & 3.12225{} & 3.12234{}  & 3.12203{} & 3.16616{}  & 3.16629{} & 3.16596{} \\ 
       &  520  & 3.00024{} & 3.00041{}  & 3.00013{} & 3.00051{}  & 3.00074{} & 3.00045{} \\
       &  920  & 2.99992{} & 3.00009{}  & 2.99980{} & 2.99986{}  & 3.00009{} & 2.99980{} \\ \hline
  32   &  500  & 6.00061{} & 6.00300{}  & 5.99995{} & 6.00000{}  & 6.00319{} & 6.00006{} \\
       &  1000 & 6.00051{} & 6.00284{}  & 5.99988{} & 5.99984{}  & 6.00298{} & 5.99990{} \\
       &  1500 & 6.00047{} & 6.00274{}  & 5.99986{} & 5.99983{}  & 6.00294{} & 5.99989{} \\
\end{tabular}
\end{ruledtabular}
\label{tbl:compareReconS}
\end{center}
\end{table} 
$S(3)$ cooling is significantly faster than $S(5)$, since it requires fewer link 
multiplications, and gives no noticeable decrease in long-term instanton stability. In 
fact, for large numbers of cooling sweeps $S(5)$ cooling appears to consistently produce 
lower values of $S/S_0$ than $S(3)$ cooling, and lower than the required integer value, 
suggesting the presence of ${\cal O}(a^6)$ discretization errors which contribute negatively. 
Over the long term these errors may destabilize instantons (as was observed 
with $S(4)$ cooling), suggesting that $S(3)$ is the safer choice for investigations 
of long-term instanton stability.\\

\section{Conclusions}
\label{sec:conclusions}
In summary, to assess the relative merits of the improvement schemes discussed in this paper, we 
compare the various cooling actions (rescaled by the single instanton action $S_0$), with 
reconstructed actions (rescaled by $S_0$) and topological charges based upon various forms of 
the improved field-strength tensor, $F_{\mu\nu}$. For studies of instantons on 
periodic (untwisted) lattices the $S(3)$ action is optimal for cooling, due to the superior 
accuracy with which it approaches integer values, its promise of long-term stability, and 
cost-effective execution. 
Five-loop improvement is optimal for the field-strength tensor, as most integer-like results 
are obtained from $Q(5)$ for the improved topological charge operator and $S_R(5)$ for the 
reconstructed action. However the 3-loop operator also produces excellent results, and is much 
less computationally demanding than the 5-loop operator. This recommends the 3-loop 
field-strength as a reasonable alternative choice to the 5-loop improved field-strength when 
computational resources are limited and improved accuracy at the order 
of a few parts in $10^4$ is not required.\\

The improved field-strength tensor has already been used in a number of investigations. For example,
the 3-loop topological charge described in this paper has been used to assess the Atiyah-Singer 
index theorem \cite{AtiyahSinger} on uncooled and cooled field 
configurations \cite{IndexOver}. The highly-improved field-strength tensor has also been used in a ``fat-link 
irrelevant clover'' (FLIC) action \cite{Adel2}, a modification of the Sheikholeslami-Wohlert improved 
quark action \cite{SheikWohlert}, where the links of irrelevant operators are fattened via APE smearing. 
Detailed investigations of the long-term behaviour of instantons 
under ${\cal O}(a^4)$-improved cooling have been performed \cite{UsNahm}, including investigations of the 
consequences of the Nahm transform \cite{Taubes,BraamBaal,Gonz-ArrPena} for the stability of $|Q|=1$ 
configurations on the 4-torus.\\

Areas of future research using the improved field-strength tensor include
adaptation to the study of glueball and hybrid meson spectra where colour-magnetic and electric fields 
explicitly provide gluonic excitations in the hadron interpolating fields \cite{Toussaint,Lacock}. 
Future investigations should also investigate dislocation thresholds to learn how coarse the lattice can be made 
before a substantial reduction in the quality of data occurs.\\

\section*{Acknowledgements}

We thank Paul Coddington and Francis Vaughan of the South Australian Centre for
Parallel Computing and the Distributed High-Performance Computing Group for support 
in the development of parallel algorithms in High Performance Fortran (HPF). 
Calculations were carried out on the Orion supercomputer at the Australian National 
Computing Facility for Lattice Gauge Theory (NCFLGT). This work was supported by the 
Australian Research Council.


\end{document}